\documentclass[prb,twocolumn,superscriptaddress,amsmath,amssymb]{revtex4-2}
\usepackage[colorlinks=true,citecolor=blue,linkcolor=blue,breaklinks=true]{hyperref}

\usepackage{color,graphicx,soul}
\usepackage{bm}
\usepackage{amsmath}
\usepackage{amssymb}
\usepackage{caption}
\usepackage[usenames,dvipsnames]{xcolor}
\usepackage{verbatim}
\usepackage{float}
\usepackage[justification=Justified]{caption}
\usepackage[defaultlines=1,all]{nowidow}

\begin{document}
	
	\title{Detection of Majorana zero modes bound to Josephson vortices in planar\\ superconductor--topological insulator--superconductor junctions}
	
	\author{Katharina Laubscher}
	
	\author{Jay D. Sau}
	
	\affiliation{Condensed Matter Theory Center and Joint Quantum Institute, Department of Physics, University of Maryland, College Park, MD 20742, USA}
	
	\date{\today}

 	\begin{abstract}
	We study signatures of Majorana zero modes (MZMs) bound to Josephson vortices in superconductor--three-dimensional topological insulator--superconductor (S--TI--S) Josephson junctions placed in a perpendicular magnetic field. First, using semiclassical analytical as well as numerical techniques, we calculate the spatially resolved supercurrent density carried by the low-energy Caroli-de Gennes-Matricon (CdGM) states in the junction. Motivated by a recent experiment~[Yue \emph{et al.}, Phys. Rev. B {\bf 109}, 094511 (2024)], we discuss if and how the presence of vortex MZMs is reflected in supercurrent measurements, showing that Fraunhofer signatures alone are not suitable to reliably detect vortex MZMs. Next, we propose two ways in which we believe supercurrent measurements could be complemented to further verify that the junction does indeed host MZMs. Explicitly, we discuss how additional Majorana signatures could be obtained by (1) mapping out the local density of states in the junction via scanning tunneling microscopy techniques, and (2) microwave spectroscopy of the spectrum of low-energy CdGM states in the junction.
	\end{abstract}
 
	\maketitle
	
	\section{Introduction}
	
	Superconductor--topological insulator (S--TI) hybrids have been proposed as a promising platform for the realization and manipulation of Majorana zero modes (MZMs). Following the seminal works by Fu and Kane~\cite{Fu2008,Fu2009}, various different ways to realize MZMs bound to the ends of proximitized TI nanowires~\cite{Cook2011,Cook2012}, magnetic vortices in planar S--TI heterostructures~\cite{Hosur2011,Ioselevich2011,Chiu2011}, or Josephson vortices in planar S--TI--S junctions~\cite{Grosfeld2011,Potter2013} have been suggested. In this work, we focus on the latter approach, which is based on the prediction that planar S--TI--S junctions host counterpropagating Majorana modes if the phase difference between the two superconductors is equal to $\pi$~\cite{Fu2008}. If, additionally, a perpendicular magnetic field is applied, the phase difference across the junction becomes spatially nonuniform, and Josephson vortices binding localized MZMs emerge whenever the local phase difference is an odd multiple of $\pi$~\cite{Grosfeld2011}. These MZMs can in principle be moved in a controlled way if the corresponding vortices are moved, e.g., by varying the magnetic field or by driving a current through the junction. Here, the planar device geometry provides additional flexibility to perform braiding operations~\cite{Fu2008,Fu2009,Grosfeld2011,Stern2019}. In this spirit, extended S--TI--S Josephson junction networks have been proposed as a powerful and versatile platform for MZM-based topological quantum computation~\cite{Hedge2020}.
	
	A number of experimental works have reported the fabrication of S--TI--S Josephson junctions as well as their characterization via supercurrent measurements~\cite{Williams2012,Qu2012,Veldhorst2012,Cho2013,Sochnikov2013,Kurter2014,Kurter2015,Sochnikov2015,Stehno2016,Kayyalha2020}. Some of these junctions were shown to exhibit non-sinusoidal (skewed) current-phase relations (CPRs), indicating the presence of high-transmittance subgap Andreev bound states (ABSs) in the junction. Furthermore, missing Shapiro steps have been interpreted as an indirect signature of the $4\pi$-periodic CPR associated with the presence of Majorana modes in the junction~\cite{Wiedenmann2016,Schuffelgen2019}. In the presence of a perpendicular magnetic field, one of the simplest and most natural ways to probe Josephson junctions is via Fraunhofer measurements. In this context, the lifting of odd nodes in the Fraunhofer pattern has recently been interpreted as a signature of localized MZMs bound to Josephson vortices in junctions based on thick TI slabs~\cite{Kurter2015,Yue2023}. This was theoretically explained by invoking a phenomenological 4$\pi$-periodic MZM contribution to the local supercurrent density at the position of the vortices~\cite{Hedge2020,Yue2023}, but microscopic calculations supporting this picture are currently lacking. Furthermore, while previous theoretical works have directly computed the CPR and Fraunhofer pattern in related but different models based on TI thin films~\cite{Potter2013} or chiral topological superconductors~\cite{Abboud2022}, it is unclear to what extent these results are applicable to the experimental situation of Refs.~\cite{Kurter2015,Yue2023}. Therefore, a microscopic analysis of the supercurrent contribution from vortex MZMs (and other low-energy states) in junctions based on thick TI slabs is of interest.

\begin{figure}[tb]
		\centering
		\includegraphics[width=1\columnwidth]{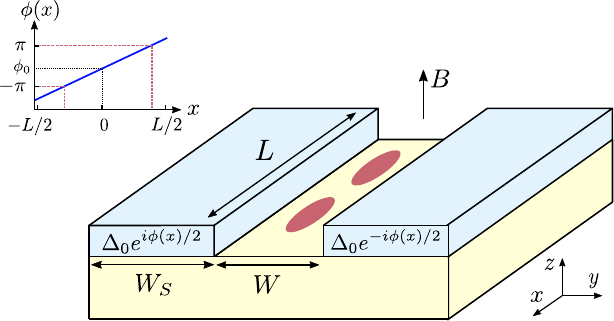}
		\caption{We consider a planar Josephson junction formed by two $s$-wave superconductors (blue) placed on the surface of a 3D TI (yellow) in the presence of a perpendicular magnetic field of strength $B$. Due to the magnetic field, the phase difference between the two superconductors varies linearly with the position along the junction (see inset). Majorana zero modes (red dots) are predicted to emerge at Josephson vortices where the superconducting phase difference is an odd multiple of $\pi$. We assume that the TI is thick enough such that only states near the top surface get proximitized, while the bottom surface remains normal.}
		\label{fig:setup}
	\end{figure}

	In the present work, we combine semiclassical analytical techniques with full numerical simulations of the two-dimensional (2D) TI surface state to study potential signatures of vortex MZMs in planar S--TI--S junctions in a perpendicular magnetic field, see Fig.~\ref{fig:setup}. First, we calculate the spatially resolved supercurrent density carried by the low-energy Caroli-de Gennes-Matricon (CdGM) states bound to a single isolated vortex. We show that isolated vortex MZMs do not significantly contribute to the local supercurrent, making it unclear whether the phenomenological model presented in Refs.~\cite{Hedge2020,Yue2023} can be used to interpret node lifting in Fraunhofer patterns~\cite{Kurter2015,Yue2023}. In fact, the node lifting found in prior theoretical calculations for related models~\cite{Potter2013,Abboud2022} originates from the interaction of vortex MZMs with the junction ends, which is not captured by the phenomenological model. In this context, however, we find that edge effects can lead to a lifting of the Fraunhofer nodes even for topologically trivial vortices that do not host MZMs. This is in addition to well-known trivial mechanisms leading to node lifting such as, e.g., an inhomogeneous magnetic field due to flux focusing effects or an inhomogeneous current distribution due to disorder or fabrication defects.
 
 Taking all of the above into account, we conclude that Fraunhofer signatures alone are not ideally suited to indicate the presence of vortex MZMs in the junction. Therefore, we next propose two ways in which we believe supercurrent measurements could be complemented to obtain a more complete understanding of the low-energy properties of S--TI--S junctions: First, the local density of states (LDOS) in the junction could be mapped out using scanning tunneling microscopy (STM) techniques, and second, microwave spectroscopy techniques could be used to obtain additional information about the spectrum of subgap CdGM states. Importantly, we focus on signatures of vortex MZMs that are located deep within the junction, excluding boundary effects that may become important as an MZM moves close to either end of the junction. This is because boundary effects can be expected to depend on details of the experimental setup, which further complicates the interpretation of potential Majorana signatures in this case.

	This paper is organized as follows. In Sec.~\ref{sec:model}, we introduce the model for the S--TI--S junction that serves as the starting point for our analytical and numerical calculations. In Sec.~\ref{sec:current_isolated}, we calculate the spatially resolved supercurrent density in the normal region of the junction in the presence of an isolated vortex MZM both analytically using a semiclassical approximation and numerically using a lattice model. Next, in Sec.~\ref{sec:LDOS}, we calculate the LDOS at the center of the junction, showing that the vortex MZMs should be visible in STM measurements as zero-energy peaks that (i) remain stable as the chemical potential is varied and (ii) move with the position of the vortices as the magnetic flux through the sample is varied. In Sec.~\ref{sec:twotone}, we further study the microwave absorption spectrum of the junction, which can be used to obtain additional information about the spectrum of subgap CdGM states. We discuss microwave signatures of isolated vortex MZMs as well as a way to extract the energy splitting between overlapping MZMs in multi-vortex configurations.

	\section{Model}
	\label{sec:model}

    We consider a planar S--TI--S Josephson junction that is formed by two $s$-wave superconductors on the surface of a 3D TI, see Fig.~\ref{fig:setup} for a schematic depiction of this setup. We assume that the TI is thick enough such that only states near the top surface get proximitized, while the bottom surface remains in the normal state. This is the situation that is believed to be relevant to the recent experimental work Ref.~\cite{Yue2023}, where no significant supercurrent was found to flow at the bottom surface of the TI. Therefore, to obtain a simple theoretical model, we only take the top surface state of the TI into account. In the additional presence of a perpendicular magnetic field of strength $B$, the S--TI--S junction can then be described by the Bogoliubov-de Gennes (BdG) Hamiltonian $H=\frac{1}{2}\int d\boldsymbol{r}\,\Psi^\dagger\mathcal{H}\Psi$ with $\Psi=(\psi_\uparrow,\psi_\downarrow,\psi_\downarrow^\dagger,-\psi_\uparrow^\dagger)^T$ and
	\begin{equation}
	\mathcal{H}=( \hbar v\boldsymbol{\sigma}\cdot\boldsymbol{\pi}-\mu)\tau_z+[\Delta(x,y)\tau_++\mathrm{H.c.}],\label{eq:H_cont}
	\end{equation} 
	where $v$ is the Fermi velocity of the surface state, $\boldsymbol{\sigma}=(\sigma_x,\sigma_y)$ is the vector of Pauli matrices acting in spin space,  $\boldsymbol{\pi}=(\pi_x,\pi_y)=(-i\partial_x-\frac{e}{\hbar} A_x\tau_z,-i\partial_y)$ is the vector of momentum with $A_x$ the vector potential in the Landau gauge, $\mu$ is the chemical potential, $\tau_{x,y,z}$ are Pauli matrices acting in particle-hole space, $\tau_\pm=(\tau_x\pm i\tau_y)/2$, and $\Delta(x,y)$ is the position-dependent proximity-induced superconducting gap. We assume that the magnetic field acts only in the normal region of the Josephson junction, such that the vector potential takes the form $A_x(y)=BW/2$ for $y<-W/2$, $-By$ for $|y|\leq W/2$, and $-BW/2$ for $y>W/2$, where $W$ is the width of the junction. Note that we neglect the Zeeman splitting induced by the magnetic field as this splitting is very small for the range of magnetic fields we are interested in ($B\sim$ a couple of mT in Ref.~\cite{Yue2023}).
    The superconducting gap is taken to be
	\begin{equation}
	\Delta(x,y)= 
	\Delta_0 e^{-i\mathrm{sgn}(y)\phi(x)/2}\Theta(|y|-W/2),\label{eq:scgap}
	\end{equation}
	where $\Delta_0>0$ is real, $\phi(x)$ is the superconducting phase difference between the two superconductors, and $\Theta$ is the Heaviside step function. In the limit where the junction length is much smaller than the Josephson penetration length, the phase becomes a linear function of the position along the junction with a slope set by the magnetic field~\cite{Yue2023,Schluck2024},
	\begin{equation}
	\phi(x)=\frac{2\pi}{L}\frac{\Phi}{\Phi_0}x+\phi_0,\label{eq:scphase}
	\end{equation}
	where $\Phi=BWL$ is the magnetic flux piercing the junction, $\Phi_0=h/2e$ is the flux quantum, $L$ is the length of the junction, and $\phi_0$ is the superconducting phase difference at the center of the junction $x=0$. (This choice of phase profile also ensures that the gauge-invariant current inside the parent superconductor vanishes.) We further assume that the magnetic field is not strong enough to induce any Abrikosov vortices inside the superconducting regions.
 
	In order to study the system numerically, we also introduce a discretized lattice model for the effective surface Hamiltonian given in Eq.~(\ref{eq:H_cont}). Assuming translational invariance for the moment, the bare TI surface state can be described by an effective Hamiltonian $\bar{H}_0=\sum_\mathbf{k}\mathbf{c}_\mathbf{k}^\dagger\,\bar{\mathcal{H}}_0\,\mathbf{c}_\mathbf{k}$ with $\mathbf{c}_\mathbf{k}=(c_{\mathbf{k},\uparrow},c_{\mathbf{k},\downarrow})^T$ and
	\begin{align}
	\bar{\mathcal{H}}_0&=\alpha[\sin(k_xa)\sigma_x+\sin(k_ya)\sigma_y]-\mu\nonumber\\&\quad-2t[\cos(k_xa)+\cos(k_ya)-2]\sigma_z,\label{eq:surface_lattice}
	\end{align}
	where we have defined $\alpha=\hbar v/a$ with $a$ the lattice spacing. In the second line of Eq.~(\ref{eq:surface_lattice}), we have added a Wilson term that breaks time-reversal symmetry away from $(k_x,k_y)=(0,0)$ in order to avoid the fermion doubling problem that arises when discretizing the 2D Dirac equation on a lattice~\cite{Vafek2014,Beenakker2023}. In real space, the BdG Hamiltonian for the planar S--TI--S junction in the presence of a magnetic field then takes the form
	\begin{align}
	\bar{H}&=\frac{1}{2}\sum_{n,m}\Big\{\Big[\frac{i\alpha}{2}\big(\boldsymbol{c}_{n+1,m}^\dagger e^{i\varphi_m\tau_z}\sigma_x\tau_z\boldsymbol{c}_{n,m}+\boldsymbol{c}_{n,m+1}^\dagger\sigma_y\tau_z\boldsymbol{c}_{n,m})\nonumber\\&\hspace{10mm}-t(\boldsymbol{c}_{n+1,m}^\dagger e^{i\varphi_m\tau_z}\sigma_z\boldsymbol{c}_{n,m}+\boldsymbol{c}_{n,m+1}^\dagger\sigma_z\boldsymbol{c}_{n,m})\nonumber\\&\hspace{10mm}+\Delta_{n,m}\boldsymbol{c}_{n,m}^\dagger\tau_+\boldsymbol{c}_{n,m}+\mathrm{H.c.}\Big]\nonumber\\&\hspace{10mm}-\mu\,\boldsymbol{c}_{n,m}^\dagger\tau_z\boldsymbol{c}_{n,m}+4t\boldsymbol{c}_{n,m}^\dagger\sigma_z\boldsymbol{c}_{n,m}\Big\}\label{eq:H_tb}
	\end{align}
	with $\boldsymbol{c}_{n,m}=(c_{n,m,\uparrow}, c_{n,m,\downarrow}, c_{n,m,\downarrow}^\dagger, -c_{n,m,\uparrow}^\dagger)^T$. Here, $\Delta_{n,m}$ is the site-dependent pairing amplitude corresponding to a discretized version of Eq.~(\ref{eq:scgap}) with a site-dependent phase $\phi_{n,m}$ given by a discretized version of Eq.~(\ref{eq:scphase}). Furthermore, the Peierls phases arising due to the magnetic field are $\varphi_m=\frac{ea}{\hbar}A_x(y)$ with $y=am$. 
    In an infinite sample, the continuum-limit low-energy spectrum of $\bar{H}$ approximates the low-energy spectrum of $H$ given in Eq.~(\ref{eq:H_cont}) as long as $\mu,\Delta_0\ll t$. At the edges of a finite sample, however, a spurious chiral edge state appears as a consequence of time-reversal symmetry breaking due to the Wilson mass term. To avoid artifacts coming from this chiral edge state in our numerical simulations, we will only use Eq.~(\ref{eq:H_tb}) to study the properties of isolated vortices deep in the junction, where edge effects can be neglected.

 \begin{figure*}[htb]
		\centering
		\includegraphics[width=1\textwidth]{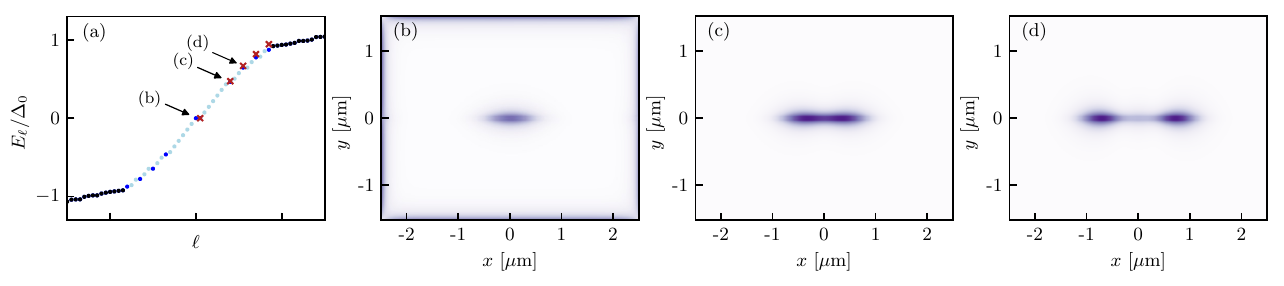}
		\caption{(a) Low-energy BdG spectrum of the discretized Hamiltonian given in Eq.~(\ref{eq:H_tb}) obtained by numerical exact diagonalization (black dots: bulk states, light blue dots: edge states, dark blue dots: CdGM states in the junction) and energies of the CdGM states in the junction obtained from the semiclassical analytical solution Eq.~(\ref{eq:energies_1d}) (red crosses; only positive energies shown). The index $\ell$ here takes integer values and enumerates the low-energy eigenstates of the numerical BdG Hamiltonian. (b)-(d) Probability density of the three states labeled by arrows in (a). The state shown in (b) is a superposition of an MZM localized at the center of the vortex and an MZM that is delocalized along the outer periphery of the superconductors. The states shown in (c) and (d) are higher-energy CdGM states bound to the vortex. We have set $v=4\times 10^5$~m/s, $\Delta_0=1.5$~meV, $\mu=0$, $W=50$~nm, $\Phi=1.3\Phi_0$, $\phi_0=\pi$, $L=5~\mu$m, and $W_S=1.5~\mu$m. In the analytical solution, this corresponds to having $\lambda\,\xi\approx0.11$ and $\tilde{v}=v$.}
		\label{fig:spectrum}
	\end{figure*}
 
 Throughout this paper, we use parameter values that roughly correspond to a Nb/Bi$_2$Se$_3$/Nb planar Josephson junction, although our results are applicable more generally. (We focus on the qualitative behavior of the system and no effort is made to quantitatively reproduce experimental results.) Unless specified otherwise, we set $v=4\times10^5$~m/s~\cite{Williams2012} and $\Delta_0=1.5$~meV~\footnote{This is an optimistic choice which simplifies our numerical simulations. A more conservative estimate for Bi$_2$Se$_3$/Nb hybrids would be $\Delta_0\sim 1$~meV~\cite{Flototto2018}.}, leading to a superconducting coherence length of $\xi=\hbar v/\Delta_0\approx 176$~nm. The chemical potential $\mu$, measured from the Dirac point, can be tuned by gating~\cite{Cho2013} and is chosen between $0$ and $10$~meV here. (Larger values of $\mu$ are possible experimentally, but detrimental to the realization of vortex MZMs as the energy gap separating the MZMs from higher-energy CdGM states in the junction decreases as $\mu$ increases~\cite{Potter2013}.) In our numerical simulations, we take the lattice constant to be $a=10$~nm and the Wilson mass is set to $t=2\alpha/5$.

\section{Supercurrent in the presence of a Josephson vortex}
\label{sec:current_isolated}
	
Our first goal is to calculate the supercurrent through a single isolated vortex located at the center $x=y=0$ of a long junction (for now, we assume $L\rightarrow\infty$). It is useful to first review the low-energy spectrum of the junction in the absence of a magnetic field, i.e., for a constant $\phi(x)\equiv \phi_0$. In this case, the momentum $k_x$ along the junction is a good quantum number. For $W=\mu=0$, it is then well-known that the Hamiltonian given in Eq.~(\ref{eq:H_cont}) admits two branches of bound state solutions with eigenvalues~\cite{Fu2008}
\begin{equation}
E(k_x)=\pm\sqrt{(\hbar vk_x)^2+\Delta_0^2\cos^2{(\phi_0/2)}}.\label{eq:abs_dispersion}
\end{equation}
These two branches correspond to a pair of dispersing subgap ABSs related to each other by particle-hole symmetry. In the special case $\phi_0=\pi$, the spectrum becomes gapless and the propagating ABSs become helical Majorana modes. At $k_x=0$, the Majorana wave functions $|\psi_{a,b}\rangle$ satisfy $|\psi_{b}\rangle\pm i|\psi_{a}\rangle\propto(\pm 1,-i,-1,\pm i)^T e^{\mp iy\mu/\hbar v-\int_0^{|y|}dy'|\Delta(y')|/\hbar v}$ up to normalization and a phase, where we have now also included a finite $\mu$ and $W\ll\xi$~\cite{Fu2008}. Small deviations from $\phi_0=\pi$ and $k_x=0$ can then be taken into account by projecting the corresponding terms in the Hamiltonian onto $|\psi_{a,b}\rangle$. This yields an effective low-energy Hamiltonian
\begin{equation}
	\mathcal{H}_{\mathrm{eff}}=-i\hbar \tilde{v}\partial_x \rho_x+m\rho_y,\label{eq:Heff1D}
\end{equation}
where $\rho_{x,y}$ are Pauli matrices acting in the space of the low-energy Majorana modes, $m=\tilde\Delta_0\cos(\phi_0/2)$ with $\tilde\Delta_0=\Delta_0/(1+W/\xi)$ is an effective mass, and $\tilde{v}$ is the renormalized velocity~\cite{Fu2008,Schluck2024}
\begin{equation}
\tilde{v}=v\,\frac{[\cos(\mu W/\hbar v)+(\Delta_0/\mu)\sin(\mu W/\hbar v)]}{1+W/\xi}\frac{\Delta_0^2}{\mu^2+\Delta_0^2}.
\end{equation}
If the small contributions $W/\xi\ll 1$ are dropped, we recover the effective Hamiltonian given in Ref.~\cite{Fu2008}.

Next, we take the external magnetic field into account, which leads to a spatial variation of $\phi(x)$ according to Eq.~(\ref{eq:scphase}). We consider the limit where the variation of $\phi(x)$ is slow, such that we can assume $\phi(x)$ to be locally independent of $x$, allowing us to calculate the supercurrent semiclassically. In the next two subsections, we will separately calculate the spatially resolved supercurrent density carried by low-energy states with $E\ll\Delta_0$ (Sec.~\ref{sec:current_close_to_vortex}) and higher-energy states with $E\lesssim\Delta_0$ (Sec.~\ref{sec:current_away_from_vortex}) in the semiclassical limit. Additionally, we will compare our analytical results to independent numerical simulations obtained from the discretized Hamiltonian given in Eq.~(\ref{eq:H_tb}).

\subsection{Current from low-energy states}
\label{sec:current_close_to_vortex}

 \begin{figure*}[hbt]
		\centering
		\includegraphics[width=1\textwidth]{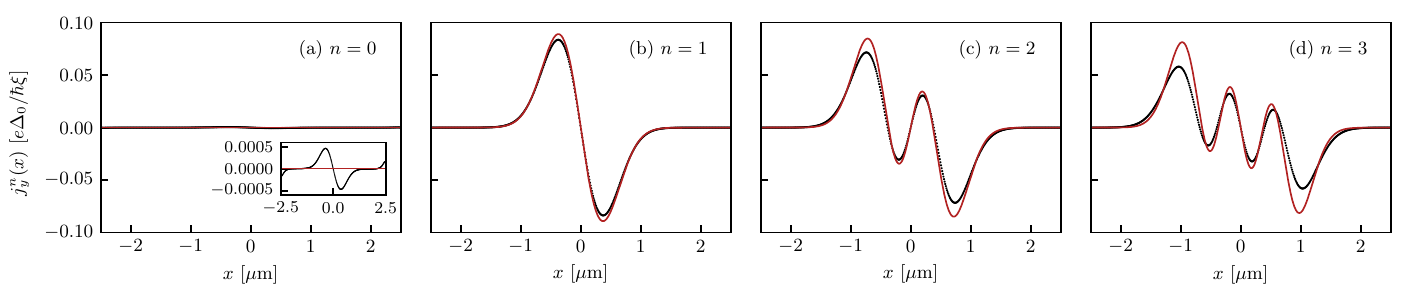}
		\caption{Contribution of individual low-lying CdGM states to the spatially resolved supercurrent density in the junction. The black dots are the numerical results obtained from the discretized Hamiltonian given in Eq.~(\ref{eq:H_tb}), and the red solid lines are obtained from the analytical solutions given in Eq.~(\ref{eq:abs_analytical}). (a) The contribution of the vortex MZM is exactly zero in the analytical solution and negligibly small ($\sim$ two orders of magnitudes smaller than the contribution from the other low-lying CdGM states) in the numerical solution, see inset. The small but finite contribution in the numerical solution comes from the hybridization between the localized vortex MZM and its delocalized partner at the outer perimeter of the junction. (b)-(d) Contributions of the CdGM states corresponding to the quantum numbers $n=1,2,3$ in Eq.~(\ref{eq:abs_analytical}), respectively. The parameters are the same as in Fig.~\ref{fig:spectrum}.}
		\label{fig:cpr_individual_ABS}
	\end{figure*}
	
	We start by focusing on the low-energy CdGM states that are localized near the center of the vortex $\phi(x)\approx \pi$. Here, we can work with the effective low-energy Hamiltonian given in Eq.~(\ref{eq:Heff1D}). The mass is now position-dependent, $m\rightarrow m(x)$, and can be expanded to $m(x)\approx -\tilde\Delta_0\left[\phi(x)-\pi\right]/2$. Assuming a slow variation of $\phi(x)$ we can further linearize $m(x)\approx -\frac{\phi'(0) \tilde\Delta_0}{2}x\equiv-\lambda\Delta_0 x$. For the phase profile given in Eq.~(\ref{eq:scphase}) we have $\lambda=\pi BW/[\Phi_0(1+W/\xi)]$ (here the linearization is actually exact), but we note that our analysis could also be adapted to other phase profiles as long as the variation of the phase is slow. The effective Hamiltonian then becomes
	\begin{equation}
	\mathcal{H}_{\mathrm{eff}}\propto\ a\rho_-+a^\dagger\rho_+,
	\end{equation}
	where we have defined $\rho_{\pm}=(\rho_x\pm i\rho_y)/2$, with  $\rho_{x,y}$ given after Eq.~(\ref{eq:Heff1D}). Furthermore, $a,a^\dagger\propto -i\hbar\tilde{v}\partial_x\mp i\lambda\Delta_0 x$ are raising and lowering operators with $[a,a^\dagger]=1$. The eigenstates of $\mathcal{H}_{\mathrm{eff}}$ can then easily be written as
	\begin{equation}
	|\psi_n\rangle=\frac{1}{\sqrt{2}}(|n\rangle\otimes|\rho_z=+1\rangle+|n-1\rangle\otimes|\rho_z=-1\rangle)\label{eq:abs_analytical}
	\end{equation}
	with eigenvalues
 \begin{equation}
 E_n=\sqrt{2 n\hbar \tilde{v}\lambda\Delta_0}\label{eq:energies_1d}
 \end{equation}
 for $n>0$. Here, we have defined the number eigenstates $|n\rangle$ satisfying $a^\dagger a|n\rangle=n|n\rangle$. The negative energy eigenstates can be obtained by applying the chiral symmetry operator $\rho_z$ so that $|\psi_{-n}\rangle\equiv\rho_z|\psi_n\rangle$ has energy $-E_n$. These in-gap ABSs with energies $\pm E_n$ are the CdGM states bound to the Josephson vortex. In addition, there is a unique MZM $|\psi_0\rangle=|0\rangle\otimes|\rho_z=+1\rangle$ with energy $E_0=0$. In Fig.~\ref{fig:spectrum}(a), we plot the first few energy levels $E_n$ and compare them to the spectrum obtained by numerical diagonalization of the discretized BdG Hamiltonian [see Eq.~(\ref{eq:H_tb})]. We set the junction width to $W=50$~nm~\cite{Schluck2024} and the chemical potential to $\mu=0$. In the numerical simulations, we further set $\Phi=1.3\Phi_0$, $\phi_0=\pi$, and $L=5~\mu$m, which corresponds to a value of $\lambda\xi\approx0.11$ in the analytical solution. (Here, the value of $\phi_0$ is chosen such that the numerical ground state energy of the system is minimized.) We find that the analytical and numerical energies of the low-energy CdGM states are in good agreement. In Fig.~\ref{fig:spectrum}(b), we additionally show the probability density of one of the zero energy states obtained from our numerical model. We see that there is one isolated MZM located at the center of the vortex as expected from our analytical analysis, while its partner (which is not captured by the analytical solution) is delocalized along the outer periphery of the superconductors~\cite{Choi2019}. The probability density for the two next CdGM states in the junction is shown in Figs.~\ref{fig:spectrum}(c-d). Note that in addition to the CdGM states, the numerical low-energy spectrum also contains edge states that propagate along the edges of the superconductors [light blue dots in Fig.~\ref{fig:spectrum}(a)]. This should be considered an artifact of our choice to only model the 2D surface state of the 3D TI. Since our model of the TI surface state given in Eq.~(\ref{eq:surface_lattice}) breaks time-reversal symmetry at the edges, our edge spectrum is closer to a TI-ferromagnet interface and therefore has gapless chiral edge states. Nevertheless, while distinct from the non-chiral surface states of a TI, the chiral edge states in our model represent the density of states outside of the junction quite well. Indeed, in setups such as the one in Ref.~\cite{Yue2023}, the experimental system is also gapless outside of the junction area due to superconductivity being absent. This is not just an experimental detail but related to the fact that Majorana zero modes must appear in pairs in any system; in the case of a single vortex, the second Majorana appears in the continuum of gapless surface states. We also note that the spectrum of low-energy states found here is similar to the low-energy spectrum found in a related work studying Josephson vortices in junctions based on 2D chiral $p$-wave superconductors~\cite{Abboud2022}. 
 
 Finally, we mention that the numerical results discussed here were obtained in the absence of disorder in the chemical potential, which is likely non-negligible in realistic samples. In the presence of disorder, we expect that the energy gap between the vortex MZM and the first trivial CdGM state is reduced compared to the clean case, similar to what was previously found for related systems~\cite{deMendon2023,Rechcinski2024}. In Appendix~\ref{app:B}, we confirm this explicitly by showing examples of CdGM spectra in the presence of spatially correlated disorder in the chemical potential. In addition, we note that band bending effects can lead to a reduced transparency of the junction, which reduces the effective superconducting gap and thereby also the spacing between CdGM states.

	With the energy spectrum at hand, we can now calculate the supercurrent carried by the low-energy CdGM states. For slowly varying $\phi(x)$, the semiclassical current operator at position $x$ is
	\begin{equation}
	j_{y}(x)=-(2e/\hbar)\partial_\phi \mathcal{H}_{\mathrm{eff}}\approx(e\tilde\Delta_0/\hbar)\rho_y.\label{eq:current_1d}
	\end{equation}
	The contribution of the $n$th low-energy CdGM state to the supercurrent density can then be obtained as $j_y^n(x)=\langle \psi_n|(j_y\otimes |x\rangle\langle x|)|\psi_n\rangle$. As can directly be seen from the form of $|\psi_0\rangle$, the isolated MZM at the center of the vortex does not contribute to the supercurrent density, $j_y^0(x)\equiv 0$, see Fig.~\ref{fig:cpr_individual_ABS}(a). 
 In particular, note that the supercurrent density at $\phi(x)\approx \pi$ from an isolated vortex MZM does not show a peak as was postulated in some previous works~\cite{Hedge2020,Yue2023}. This behavior is further confirmed by numerical results obtained from the discretized model given in Eq.~(\ref{eq:H_tb}), in which case we find a finite but negligibly small Majorana contribution to the supercurrent density ($\sim$ two orders of magnitude smaller than the contribution from the other low-lying CdGM states). Here, we have calculated the current at the center of the junction ($y=0$) using a discretized version of the current operator for our model of the TI surface state, which is given in Appendix~\ref{app:A} for completeness. The small but finite Majorana contribution in the numerical solution comes from the hybridization between the localized vortex MZM and its delocalized partner at the outer perimeter of the junction. However, even on this scale, one does not find a peak in the supercurrent density at the position of the vortex, but rather a dip--peak structure centered symmetrically around the vortex, at the center of which the supercurrent density vanishes and flips sign.
 
 The contribution of the CdGM states with $n>0$ can be obtained from Eqs.~(\ref{eq:abs_analytical}) and (\ref{eq:current_1d}) as $j_y^n(x)=(e\tilde\Delta_0/\hbar)\,\mathrm{Im}[\varphi_{n-1}(x)\varphi_n^*(x)]$, where we have introduced the wave functions $\varphi_n(x)=\langle x|n\rangle$. We plot $j_y^{n}(x)$ for $n=1,2,3$ in Figs.~\ref{fig:cpr_individual_ABS}(b-d). In addition to the semiclassical analytical results obtained via Eqs.~(\ref{eq:abs_analytical}) and (\ref{eq:current_1d}), we again also show the numerical results obtained from the discretized model given in Eq.~(\ref{eq:H_tb}), showing that the two sets of results agree well at low energies. For our numerical model, we have further checked that the edge states [light blue dots in Fig.~\ref{fig:spectrum}(a)] do not significantly contribute to the supercurrent.

 While we have focused on the spatially resolved supercurrent density so far, we also briefly comment on the total (integrated) current carried by the low-lying CdGM states. Our results in Fig.~\ref{fig:cpr_individual_ABS} show that, for a spatially isolated vortex, all low-energy CdGM states (including the vortex MZM) yield negligible contributions to the total current since the CdGM contributions $j_y^{n}(x)$ to the supercurrent density are antisymmetric about the vortex center. This is consistent with the fact that the net current around an isolated vortex, even including CdGM contributions, is zero as required by current conservation. The case of non-isolated vortices and, in particular, the importance of boundary effects is discussed in Sec.~\ref{sec:fraunhofer}.

	\subsection{Current from higher-energy states}
	\label{sec:current_away_from_vortex}
 
\begin{figure}[tb]
		\centering
		\includegraphics[width=0.9\columnwidth]{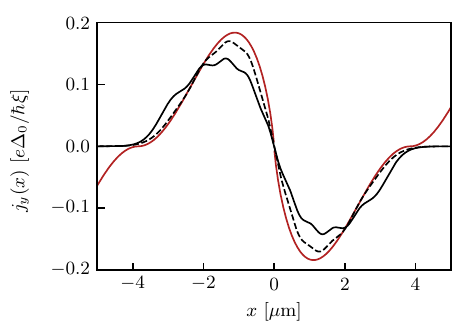}
		\caption{Total current density carried by the subgap CdGM states in the junction. The black dashed (solid) line corresponds to the numerical result for a junction of width $W=10$~nm ($W=50$~nm), where we have added up the supercurrent contributions of all subgap CdGM states as obtained by exact diagonalization of Eq.~(\ref{eq:H_tb}). The red solid line corresponds to the approximate analytical solution given in Eq.~(\ref{eq:cprTot}). Except at the junction ends, Eq.~(\ref{eq:cprTot}) approximates the numerical result well for narrow junctions. In particular, the total current density does indeed vanish and flip sign at the center of the vortex. The other parameters are the same as in Fig.~\ref{fig:spectrum}, except that we have considered a larger system with $L=10~\mu$m to increase the number of subgap states in the junction.}
		\label{fig:cpr_full}
	\end{figure}

	It is also possible to obtain an approximate semiclassical expression for the supercurrent carried by the higher-energy subgap states. For this, we start from the dispersion relation Eq.~(\ref{eq:abs_dispersion}), where the momentum $k_x$ satisfies $|k_x|<\Delta_0|\sin{(\phi(x)/2)}|/\hbar v$ such that the energy is bound to be below $\Delta_0$. The current from the bound state is then given by a derivative~\cite{Beenakker1991}
	\begin{equation}
	j_y(k_x,\phi(x))=-\frac{2e}{\hbar}\partial_\phi E(k_x,\phi(x))=\frac{2e}{\hbar}\frac{\Delta_0^2\sin{\phi(x)}}{4E(k_x,\phi(x))}.
	\end{equation}
	Considering the continuum limit of a large number of excitations, the total current density can be obtained as 
	\begin{align}
	j_y(x)&=\frac{2e}{\hbar}\int_0^{\Delta_0|\sin{(\phi(x)/2)}|/\hbar v} \frac{dk_x}{2\pi}\frac{\Delta_0^2\sin{\phi(x)}}{2E(k_x,\phi(x))}
	\nonumber\\&=-\frac{2e}{\hbar}\frac{\Delta_0^2\sin{\phi(x)}}{8\pi\hbar v}\textrm{ln}\left(\frac{1-|\sin{(\phi(x)/2)|}}{1+|\sin{(\phi(x)/2)|}}\right).\label{eq:cprTot}
	\end{align}
 Note that in the above we have ignored the contribution from continuum states, which is small in the limit $W\ll\xi$~\cite{Beenakker1991b}.
 In Fig.~\ref{fig:cpr_full}, we plot the current density from the expression given in Eq.~(\ref{eq:cprTot}) for the phase configuration $\Phi=1.3\Phi_0$, $\phi_0=\pi$ and compare it to the current density obtained from our numerical model for the same phase configuration and two different junction widths $W=10$~nm, $50$~nm. In the numerical model, we have added up the supercurrent contributions of all subgap CdGM states as obtained by exact diagonalization of Eq.~(\ref{eq:H_tb}). Away from the junction ends, we find that Eq.~(\ref{eq:cprTot}) approximates the numerical result well for very narrow junctions, while the agreement becomes more qualitative as the width of the junction increases. This is expected since the dispersion relation Eq.~(\ref{eq:abs_dispersion}) is only exact in the limit $W=\mu=0$. The discrepancy at the junction ends comes from the fact that the analytical solution was derived assuming a junction of infinite length. Nevertheless, our main qualitative finding is that the current density vanishes at the center of the vortex at $x=0$ in agreement with our previous analysis of the low-energy bound states near the vortex. However, in order to correctly capture the contribution of these low-energy states around $\phi(x)\sim \pi$, the more accurate solution presented in the previous subsection should be used since the discretization of the states becomes important at these low energies.

 \subsection{Total current through the junction}\label{sec:fraunhofer}
 
 Fraunhofer pattern measurements~\cite{Williams2012,Kurter2015} depend on the total CPR rather than the spatially resolved supercurrent density shown in Fig.~\ref{fig:cpr_full}. Within our numerical model, the total supercurrent $I(\phi_0,\gamma=\Phi/\Phi_0)$ can in principle be obtained by integrating the supercurrent density over position, $I(\phi_0,\gamma)=\int_{-L/2}^{L/2}dx\, j_y(x)$. The critical current, obtained by maximizing $I(\phi_0,\gamma)$ over $\phi_0$, then follows a Fraunhofer pattern similar to the one obtained in Ref.~\cite{Abboud2022} for a related model based on chiral topological superconductors. However, since our numerical model is still considerably oversimplified compared to the actual experimental situation of Refs.~\cite{Williams2012,Kurter2015,Yue2023}, we do not make any effort here to directly relate our numerical results to the experimentally observed Fraunhofer patterns in these works.
 
 Instead, we only comment on the main qualitative features of the CPR, and, in particular, the similarities and differences between our numerical CPR and the CPR estimated in Refs.~\cite{Hedge2020,Yue2023} based on the conventional theory of extended Josephson junctions~\cite{Tinkham2004,Barone1982}. In this framework, the total current is obtained by integrating the local current $I_0(\phi(x))$ along the length of the junction,
 \begin{align}
&I(\phi_0,\gamma)=\int_{-L/2}^{L/2}dx\, J(x) I_0(\phi(x)).\label{eq:local_cpr}
\end{align}
Here, $\phi(x)$ is given in Eq.~(\ref{eq:scphase}), and we have introduced the relative Josephson coupling strength $J(x)$ along the length of the junction, which is set to unity in the context of Refs.~\cite{Hedge2020,Yue2023}. Importantly, note that writing the local current as $I_0(\phi(x))$ implies that the spatially resolved current density is translated in $x$ by $\delta \phi_0 L/2\pi \gamma$ when 
 the phase is changed by $\delta\phi_0$. Numerically, however, this turns out to only be true away from the ends of the junction. Thus, the CPR obtained using Eq.~(\ref{eq:local_cpr}) leads only to an approximate CPR for extended Josephson junctions. In particular, boundary effects are not captured in the standard description with $J(x)=1$.

\begin{figure}[tb]
	\centering
	\includegraphics[width=0.9\columnwidth]{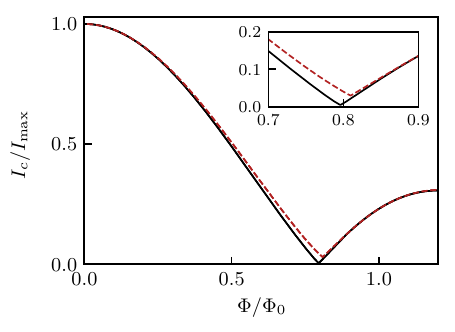}
		\caption{Normalized critical current as a function of flux calculated from Eqs.~(\ref{eq:local_cpr}) and (\ref{eq:cpr_skewed}) assuming a modified Josephson coupling at the junction ends with $J_1=0.15L$, see main text. Black solid (red dashed) line: $K=0.6$ ($K=0.95$). We find that the first node in the Fraunhofer pattern stays at zero for the more sinusoidal CPR, but gets lifted for the highly skewed CPR due to edge effects at the junction ends. The inset shows a zoom into the region around the Fraunhofer node.}
	\label{fig:fraunhofer}
\end{figure}

It turns out that exactly such boundary effects are important to explain the node lifting found in previous microscopic calculations for related models based on TI thin films~\cite{Potter2013} or chiral topological superconductors~\cite{Abboud2022}. In these models, node lifting has been shown to occur due to a nonzero MZM contribution to the total current that arises when a vortex MZM approaches the end of the junction, where it then hybridizes with an additional Majorana mode that is bound to the bottom surface of the TI~\cite{Potter2013} or delocalized along the periphery of the junction~\cite{Abboud2022}. However, as we elaborate below, even the interaction of trivial vortices with the junction ends could give rise to a similar lifting of nodes, such that node lifting in Fraunhofer patterns cannot be reliably associated with the topological superconducting character of vortices. To see this, we first note that the conventional Fraunhofer relation Eq.~(\ref{eq:local_cpr}) cannot lead to node lifting with $J(x)=1$. In this case, the Josephson current $I(\phi_0,\gamma=1)$ vanishes since $\phi(x)$ winds over the entire period of $2\pi$, leading to a node in the Fraunhofer pattern. However, in the case where $\phi(x)=\pi$ localizes a Josephson vortex (topological or trivial), a local energy minimum favoring a specific value of $\phi_0$ may be created from an effective interaction of the vortex with the junction ends. This can give rise to a change in the relative Josephson coupling at the ends of the junction, which we simplistically model as $J(x)=1 +J_1 [\delta(x-L/2)+\delta(x+L/2)]$. Here, $J_1$ is the relative strength of the modified Josephson coupling at the ends of the junction. The critical current continues to vanish at nodes for sinusoidal $I_0(\phi)$ even in the presence of a finite $J_1$. However, for highly transparent junctions, the CPR is modified to~\cite{Spanton2017}
\begin{align}
    &I_0(\phi)=J_0\,K^2 \sin{\phi}/\sqrt{1-K^2 \sin^2{(\phi/2)}},\label{eq:cpr_skewed}
\end{align}
where $J_0$ is the Josephson coupling at the center of the junction and $K$ is related to the transparency of the junction.
Computing the critical current as a function of flux as shown in Fig.~\ref{fig:fraunhofer} shows that in this case the nodes of the critical current are lifted similar to those previously associated with MZMs~\cite{Potter2013,Abboud2022}. However, the node lifting found here is simply a consequence of edge effects arising at the junction ends and can occur regardless of whether the vortices are topological or not. Furthermore, this mechanism is quite apart from reflection symmetry breaking of the junction~\cite{Kurter2015}, which is another, even simpler way to lift nodes in the Fraunhofer pattern. Since all of these mechanisms could lead to node lifting, it is complicated to separate vortex MZM-based mechanisms from trivial ones.

 Taking all of the above into account, we believe that Fraunhofer signatures are not ideally suited to indicate the presence of vortex MZMs in the junction. Nevertheless, supercurrent-based signatures could be complemented by other indicators in order to obtain more conclusive evidence of the presence or absence of MZMs in the junction. In the following two sections, we present two ways in which such additional information could be obtained.

\section{Local density of states}
\label{sec:LDOS}

\begin{figure}[tb]
	\centering
	\includegraphics[width=0.5\textwidth]{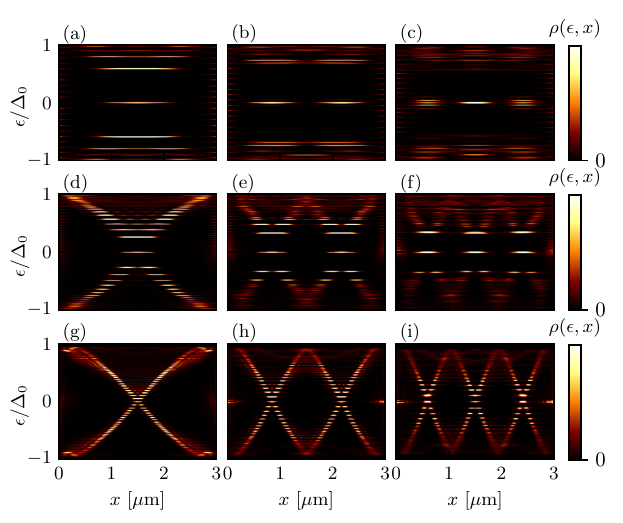}
		\caption{Local density of states (arbitrary units) at the center of the junction obtained by numerically evaluating Eq.~(\ref{eq:ldos}) for different phase configurations (a,d,g) $\Phi=1.3\Phi_0$, $\phi_0=\pi$, (b,e,h) $\Phi=2.3\Phi_0$, $\phi_0=0$, and (c,f,i) $\Phi=3.3\Phi_0$, $\phi_0=\pi$, and different values of chemical potential (a-c) $\mu=0$, (d-f) $\mu=3$~meV, and (g-i) $\mu=10$~meV. The vortex MZMs manifest as zero-energy peaks in the LDOS that (i) move with the position of the vortex as the magnetic flux through the sample changes and (ii) remain stable as the chemical potential is varied. The other parameters are $v=4\times10^5$~m/s, $\Delta_0=1.5$~meV, $W=50$~nm, $L=3~\mu$m, and $W_S=1.5~\mu$m.}
	\label{fig:ldos_mu}
\end{figure}

To obtain further information about the low-energy spectrum of the junction, the local density of states (LDOS) in the normal section could be mapped out using STM techniques. 
Such LDOS measurements in Josephson junctions have previously been proposed~\cite{Tanaka1996} to characterize the nature of pairing in the adjacent superconductor.
In the context of Josephson vortices, a recent experiment~\cite{Roditchev2015} reports the successful STM imaging of vortex cores in Pb-based SNS junctions, demonstrating that such measurements are feasible in principle. Within our numerical model [see Eq.~(\ref{eq:H_tb})], we can calculate the LDOS as
\begin{equation}
\rho(\epsilon,\boldsymbol{r})=-\sum_{\sigma,E_k\geq 0} [|u^k_\sigma(\boldsymbol{r})|^2f'(\epsilon-E_k)+|v^k_\sigma(\boldsymbol{r})|^2f'(\epsilon+E_k)],\label{eq:ldos}
\end{equation}
%
where $u_\sigma^k(\boldsymbol{r})$ [$v_\sigma^k(\boldsymbol{r})$] are the electron (hole) components of the BdG eigenstate with energy $E_k$. Furthermore, $f(\epsilon)=1/[\exp(\epsilon/k_BT)+1]$ is the Fermi-Dirac distribution function at temperature $T$, and $f'(\epsilon)=\partial f(\epsilon)/\partial\epsilon$. In Fig.~\ref{fig:ldos_mu}, we display the LDOS $\rho(\epsilon,\boldsymbol{r}=(x,0))\equiv \rho(\epsilon,x)$ along a cut through the junction center at $y=0$. The temperature is set to $T=50~$mK. We show results for three different phase configurations corresponding to different values of flux $\Phi=1.3\Phi_0,$ $2.3\Phi_0$, $3.3\Phi_0$, and three different chemical potentials $\mu=0$, $3$~meV, $10$~meV. We observe that the energy spacing between the CdGM states increases (decreases) as the flux through the sample (the chemical potential) increases. Both of these effects are expected from Eq.~(\ref{eq:energies_1d}) since $\lambda$ increases as the flux increases, while the renormalized velocity $\tilde{v}$ decreases as the chemical potential increases. Additionally, we find that the localization length of the vortex MZMs along the $x$ direction decreases as the chemical potential increases. In all panels, the localization length of the MZMs is much shorter than the intervortex distance, such that no hybridization between neighboring MZMs is observed. [However, in Fig.~\ref{fig:ldos_mu}(c), the outer two MZMs hybridize with the continuum at the sample edge and therefore split away from zero energy.] In general, spatially isolated vortex MZMs should be visible in STM measurements as zero-energy peaks in the LDOS that (i) move with the position of the vortex as the magnetic flux through the sample changes and (ii) remain stable as the chemical potential is varied. While (near-)zero-energy states can, under certain conditions, also appear in trivial junctions, we note that these trivial states are expected to split away from zero as the chemical potential is varied.

One potential drawback of STM measurements is their limited energy resolution, which might make it impossible to resolve individual MZMs immersed in a background of low-energy CdGM states. This limitation becomes more severe for small magnetic fields and large chemical potentials since the energy spacing between the vortex MZMs and the higher-energy CdGM states decreases with increasing $\mu$ and decreasing $\Phi$ (see above). For our choice of system parameters, the energy splitting between the vortex MZMs and the next CdGM states is very large ($\sim 880~\mu$eV) in Fig.~\ref{fig:ldos_mu}(a) and remains sizable ($\sim 80~\mu$eV) even in Fig.~\ref{fig:ldos_mu}(g), such that the MZMs should remain resolvable by STM even for chemical potentials exceeding $10$~meV. However, this is partly a consequence of our optimistic choice of $\Delta_0=1.5$~meV for the proximity-induced gap. (A more conservative choice for Bi$_2$Se$_3$/Nb hybrids would be $\Delta_0\sim 1$~meV~\cite{Flototto2018}.) Furthermore, the gaps become significantly smaller if---as in many current devices---Al rather than Nb is used as the superconductor, in which case the induced gap is only $\Delta_0\sim 150~\mu$eV to begin with. Therefore, it would be desirable to have additional detection schemes available that can resolve the subgap spectrum at smaller energy scales. In the next section, we will discuss microwave spectroscopy of the junction as one possible way forward.

We note that instead of using STM techniques, it is also possible to obtain information about the LDOS by tunneling spectroscopy measurements. In this case, if the MZMs are moved along the junction by tuning the global phase offset $\phi_0$ (e.g., via a flux loop), the tunneling conductance maps should show a zero-bias peak whenever an MZM passes by a lead. A very recent experiment~\cite{Schluck2024} reports the successful observation of such signatures. (However, due to limitations in energy resolution, the energy gap between the MZMs and the lowest-energy CdGM states was not resolved here.)

\section{Microwave spectroscopy}
\label{sec:twotone}

Additional information about the spectrum of low-energy CdGM states can be obtained by microwave spectroscopy techniques. Here, microwave irradiation induces parity-conserving transitions between ABSs that can be detected in a circuit quantum electrodynamics (cQED) setup. Such techniques have successfully been employed to probe the ABS spectra of nanowire Josephson junctions~\cite{Janvier2015,VanWoerkom2017,Hays2018,Fatemi2022} and, more recently, also of planar Josephson junctions based on proximitized semiconductor two-dimensional electron gases (2DEGs)~\cite{Chidambaram2022,Hinderling2023,Hinderling2024,Elfeky2024}. In the following, we discuss how vortex MZMs in planar S--TI--S junctions could be detected using microwave spectroscopy techniques, extending and adapting some of the ideas that have previously been  brought forward in the context of other Majorana platforms such as semiconductor nanowires~\cite{Virtanen2013,Ginossar2014,Vayrynen2015,Dmytruk2015,Peng2016,Dmytruk2016,Heck2017,Trif2018,Smith2020,Kurilovich2021,Dmytruk2023,Dmytruk2023b}, semiconductor 2DEGs~\cite{Pekerten2024}, TI nanowires~\cite{Yavilberg2019}, and iron-based superconductors~\cite{Ren2024}.

\begin{figure}[tb]
		\centering
		\includegraphics[width=\columnwidth]{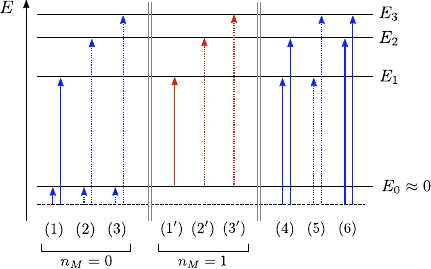}
		\caption{Schematic depiction of low-energy microwave-induced transitions between the four lowest-energy CdGM states in a junction with a single isolated Josephson vortex. The black dashed line denotes the superconducting ground state, the energy level $E_0\approx 0$ corresponds to the Majorana state, and the energy levels $E_1,...,E_3$ denote the three next CdGM states. The blue processes labeled (1)--(6) denote even transitions, where a pair of quasiparticles is excited above the superconducting ground state, while the red processes labeled (1')--(3') denote odd transitions, where a single quasiparticle is excited to a higher energy level. Processes (1)--(3) [(1')--(3')] require an unoccupied [occupied] Majorana state $n_M=0$ [$n_M=1$]. For a well-isolated vortex MZM at $E_0\approx 0$, the observed transition frequencies are independent of the parity of the Majorana state since $E_k+E_0\approx E_k-E_0$. The solid (dashed) arrows denote the processes that are dominant (suppressed) in the regime where Eq.~(\ref{eq:abs_analytical}) is applicable.}
		\label{fig:abs_transitions}
	\end{figure}

The allowed microwave-induced transitions between ABSs can be found by studying the frequency-dependent dissipative (real) part of the junction admittance $\mathrm{Re}\,Y(\omega)$, which is related to the microwave absorption rate $W$ of the junction as $W\propto\omega\mathrm{Re}\,Y(\omega)$ for $\omega>0$. Throughout this section, we consider the case of zero temperature for simplicity, although our results could readily be generalized to the finite-temperature case. In linear response, and setting $\hbar=1$ hereinafter, the real part of the junction admittance can be written as~\cite{Heck2017}
\begin{align}
\mathrm{Re}\,Y(\omega)&=\frac{\pi}{\omega}\sum_{E_k>E_l\geq 0}\Big[|j_{k,l}|^2(n_l-n_k)\delta(E_k-E_l-\omega)\nonumber\\&+|j_{k,-l}|^2(1-n_l-n_k)\delta(E_k+E_l-\omega)\Big],\label{eq:admittance}
\end{align}
where the off-diagonal matrix elements of the current operator are given by $j_{k,l}=\langle\psi_k|j_y|\psi_l\rangle$. Here, $\psi_{k}$ denotes an eigenstate of the junction Hamiltonian with energy $E_{k}\geq 0$, and $\psi_{-k}$ denotes the particle-hole partner of $\psi_k$ at energy $-E_k$. Furthermore, $n_k\in\{0,1\}$ denotes the occupation number of the state corresponding to $\psi_k$. The terms $\propto |j_{k,-l}|^2$ correspond to even transitions, where a pair of quasiparticles is excited above the superconducting ground state, while the terms $\propto |j_{k,l}|^2$ correspond to odd transitions, where a single quasiparticle is excited to a higher energy level. A schematic depiction of allowed even and odd transitions involving the four lowest-energy CdGM states in a junction with a single isolated Josephson vortex is shown in Fig.~\ref{fig:abs_transitions}. The blue processes labeled (1)--(6) denote even transitions, while the red processes labeled (1')--(3') denote odd transitions. Processes (1)--(3) [(1')--(3')] require an unoccupied [occupied] Majorana state $n_0\equiv n_M=0$ [$n_M=1$]. For a spatially isolated vortex MZM at $E_0\approx 0$, the transition frequencies are independent of the parity of the Majorana state since $E_k+E_0\approx E_k-E_0$. Therefore, we expect to observe only a single set of absorption lines even if the measurement is repeated over times longer than the typical quasiparticle poisoning time. Such `parity-independent' absorption spectra that remain stable as a function of, e.g., chemical potential or magnetic field are an indicator for isolated MZMs in the junction.

\begin{figure}[tb]
		\centering
		\includegraphics[width=0.9\columnwidth]{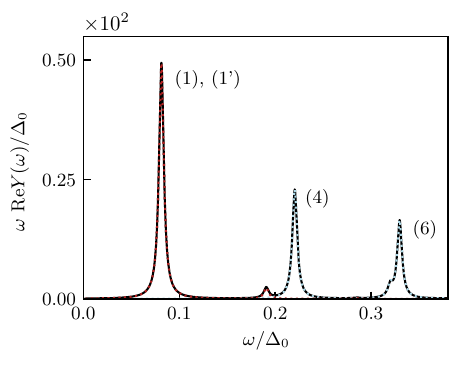}
		\caption{Real part of the junction admittance for an S--TI--S junction with one isolated Josephson vortex. The black solid  line shows the absorption spectrum obtained by numerically evaluating Eq.~(\ref{eq:admittance}) with a finite broadening $\Gamma=3~\mu$eV in the case $n_M=0$ (only even transitions), while the colored dashed lines show the absorption spectrum in the case $n_M=1$ (red: odd transitions, blue: even transitions). We find that the transition frequencies for $n_M=0,1$ are identical as expected for an isolated MZM. The labels (1), (1'), (4), and (6) identify the corresponding transitions in Fig.~\ref{fig:abs_transitions}. We have set $v=4\times10^5$~m/s, $\Delta_0=1$~meV, $W=50$~nm, $L=4~\mu$m, $W_S=2.25~\mu$m, $\mu=7$~meV, $\Phi=1.3\Phi_0$, and $\phi_0=\pi$. For simplicity, we have further set $e=\hbar=1$.}
		\label{fig:abs_absorption_lines}
	\end{figure}

We can obtain intuition about the allowed transitions by evaluating the off-diagonal matrix elements of the current operator given in Eq.~(\ref{eq:current_1d}) for the analytical low-energy solutions given in Eq.~(\ref{eq:abs_analytical}). In this case, we find that the only nonzero matrix elements entering Eq.~(\ref{eq:admittance}) are of the form $|j_{n+1,\pm n}|^2$. In Fig.~\ref{fig:abs_transitions}, these dominant transitions are shown as solid arrows, while all other transitions are shown as dashed arrows to indicate that they are suppressed in the regime where Eq.~(\ref{eq:abs_analytical}) is applicable. At this point, it is also worth noting that our analytical solution is not able to capture the difference between even and odd parity states as it captures only a single MZM. 

To go beyond our analytical low-energy solution, we can evaluate Eq.~(\ref{eq:admittance}) numerically using the eigenenergies and eigenstates of Eq.~(\ref{eq:H_tb}) and the discretized current operator given in Appendix~\ref{app:A}. Here, we include a broadening of the absorption lines due to, e.g., finite dissipation, which we include phenomenologically by replacing the delta functions in Eq.~(\ref{eq:admittance}) by a Lorentzian, $\delta(\omega)\rightarrow \Gamma/[\pi (\Gamma^2+\omega^2)]$, where $\Gamma$ determines the width of the absorption lines. In Fig.~\ref{fig:abs_absorption_lines}, we display the absorption spectra for $n_M=0,1$ at a fixed value of flux $\Phi=1.3\Phi_0$ for a junction with a single isolated Josephson vortex. As expected in the presence of an isolated MZM, we find that the absorption spectra for the two Majorana parities $n_M=0,1$ look indistinguishable. Importantly, the absorption lines remain parity-independent for an extended range of magnetic fields and chemical potentials, whereas parity-independent absorption lines due to accidental trivial zero-energy states would typically only occur for specific fine-tuned values of the system parameters.

\begin{figure*}[bt]
		\centering
		\includegraphics[width=\textwidth]{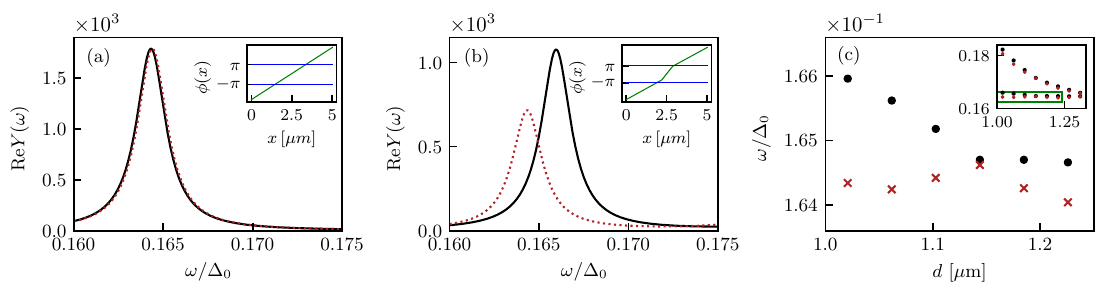}
		\caption{Real part of the junction admittance $\mathrm{Re}\,Y(\omega)$ for a junction with two Josephson vortices. The energy windows in (a) and (b) are chosen such that only the lowest-energy microwave-induced transition is visible. This transition is an even (odd) transition for $n_M=0$ ($n_M=1$). (a) Linear phase profile with a constant slope set by $\Phi=2.6\Phi_0$ (see inset). Since the overlap between the two vortex Majoranas is negligible here, the transition frequencies for $n_M=0$ (black solid line) and $n_M=1$ (red dashed line) are practically identical, leading to a single peak in $\mathrm{Re}\,Y(\omega)$. (b) Phase profile where the slope between the two vortices is locally increased by, e.g., an additional current pulse (see inset). Due to the increased overlap between the two vortex Majoranas, we find that the transition frequencies for $n_M=0$ (black solid line) and $n_M=1$ (red dashed line) differ. (c) Transition frequencies [i.e., peaks in $\mathrm{Re}\,Y(\omega)$] as a function of the intervortex distance $d$. The inset shows the two lowest-energy peaks in the junction admittance for $n_M=0$ (black dots) and $n_M=1$ (red crosses), respectively, as a function of the intervortex distance. The main figure corresponds to a zoom into the region framed by the green box in the inset. We have set $v=4\times10^5$~m/s, $\Delta_0=1$~meV, $W=50$~nm, $L=5~\mu$m, $W_S=2.25~\mu$m, $\mu=5$~meV, $\phi_0=\pi/10$, and $\Gamma=1~\mu$eV in all panels. For simplicity, we have further set $e=\hbar=1$.}
		\label{fig:mw_two_vortices}
\end{figure*}

Microwave spectroscopy techniques can additionally be used to probe the energy splitting between a pair of overlapping vortex MZMs. As an illustration, we consider the case $\Phi=2.6\Phi_0$, such that two vortices are present in the junction. In Fig.~\ref{fig:mw_two_vortices}(a), we show the absorption peak in the real part of the junction admittance resulting from transitions involving the vortex MZM state and the first CdGM state in the junction. The black solid (red dashed) line corresponds to the even (odd) transition for $n_M=0$ ($n_M=1$). Since the junction is much longer than the localization length of the MZMs, the MZMs are well separated and the even and odd transition frequencies are practically identical. In order to decrease the distance between the two vortices while keeping both vortices far away from the junction ends, one can consider a modified setup where a current pulse is applied to a narrow loop of wire crossing the junction~\cite{Hedge2020}, such that the slope of the superconducting phase difference $\phi(x)$ locally increases between the two vortices, see the inset of Fig.~\ref{fig:mw_two_vortices}(b) for a schematic depiction. This decreases the distance between the two vortices, which in turn increases the energy splitting between the states with $n_M=0,1$. This energy splitting can be observed as a splitting between the even and odd transition frequencies, see Fig.~\ref{fig:mw_two_vortices}(b). Finally, if the splitting between the even and odd transition frequencies is tracked as a function of the intervortex distance (controlled, e.g., by the current through the additional wire loop~\cite{Hedge2020}), one observes an oscillating behavior that is related to the oscillating energy splitting between overlapping MZMs, see Fig.~\ref{fig:mw_two_vortices}(c).

The microwave absorption signatures discussed in this section can, e.g., be accessed in two-tone spectroscopy measurements, which essentially probe the real part of the admittance $\mathrm{Re}\,Y(\omega)$ at different frequencies. Since such techniques can resolve energy differences at the sub-$\mu$eV scale, it should be possible to detect even small parity-dependent changes in the allowed transition frequencies as a function of, e.g., the intervortex distance. Alternatively, the imaginary part of the junction admittance $\mathrm{Im}\,Y(\omega)$ could be accessed in a cQED setup by measuring the frequency shift of the resonator due to the coupling to the junction. Again, the shift of the resonator frequency will be parity-independent for well-separated MZMs and become parity-dependent as the MZMs are brought closer together. Since the successful coupling of a microwave cavity to an S--TI--S junction has already been reported in a recent experimental work~\cite{Schmitt2022}, we believe that the microwave spectroscopy measurements proposed here may be within experimental reach.

\section{Conclusions}
\label{sec:conclusions}

Motivated by a recent experiment~\cite{Yue2023}, we have studied signatures of MZMs bound to Josephson vortices in planar S--TI--S junctions placed in a perpendicular magnetic field. We have calculated the supercurrent density carried by the low-energy CdGM states in the junction both analytically and numerically, showing that the MZM contribution to the supercurrent is vanishingly small for isolated vortex MZMs. In particular, isolated vortex MZMs do not lead to a peak in the supercurrent density at the position of the vortex, making it unclear whether the phenomenological models presented in Refs.~\cite{Hedge2020,Yue2023} can be used to interpret node lifting in Fraunhofer patterns. This does not contradict related previous theoretical works that predicted a lifting of Fraunhofer nodes in topological Josephson junctions based on TI thin films~\cite{Potter2013} or chiral topological superconductors~\cite{Abboud2022} since the node lifting discussed in these works arises from the interaction of MZMs with the edge of the device, which is an effect that is not captured by the phenomenological model. However, we show that a rather generic model of end effects in extended Josephson junctions can produce node lifting in Fraunhofer patterns even for non-topological vortices. As such, node lifting in Fraunhofer patterns cannot be reliably associated with the topological superconducting character of vortices.

We have proposed two additional measurements that could be used to further verify whether the junction hosts MZMs or not. First, STM measurements of the LDOS should reveal robust zero-energy peaks that move with the position of the vortices as the magnetic field is varied. Second, microwave spectroscopy techniques could be used to obtain additional information about the subgap spectrum of the junction. Here, isolated MZMs are expected to lead to robust parity-independent absorption lines that remain stable over long times (longer than the typical quasiparticle poisoning time), whereas an oscillating splitting between even and odd transition frequencies should become visible when the distance between two neighboring vortex MZMs is decreased. While none of the measurements proposed here will, in isolation, provide a `smoking-gun' signature for vortex MZMs, the simultaneous observation of several of the independent characteristic properties of the subgap CdGM spectrum discussed here will increasingly constrain the set of trivial explanations that could potentially reproduce the same signatures.

\section*{Acknowledgments}
This work is supported by the Laboratory for Physical Sciences through the Condensed Matter Theory Center. JS acknowledges support 
from the Joint Quantum Institute at the University of Maryland. JS thanks Victor Yakovenko for valuable discussions that motivated this work. We also thank Smitha Vishveshwara for explaining the subtleties of the phenomenological model for the Fraunhofer patterns.

\begin{figure*}[tb]
		\centering
		\includegraphics[width=1\textwidth]{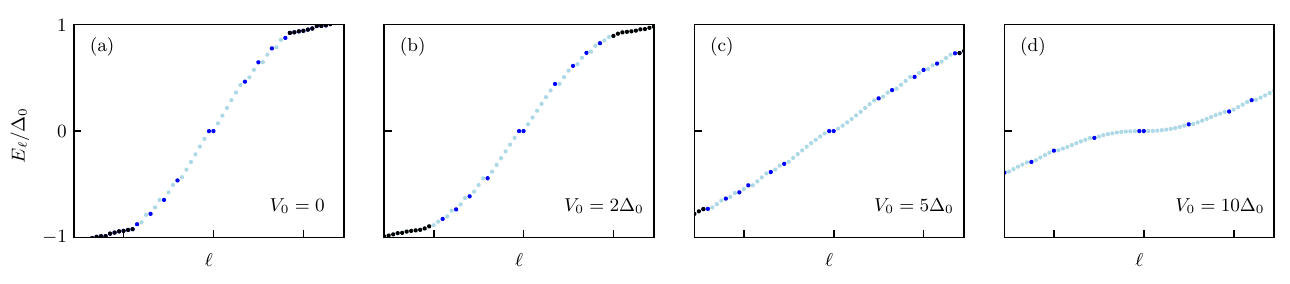}
		\caption{Low-energy BdG spectrum of the discretized Hamiltonian given in Eq.~(\ref{eq:H_tb}) obtained by numerical exact diagonalization (black dots: bulk states, light blue dots: edge states, dark blue dots: CdGM states in the junction) in the additional presence of spatially correlated disorder in the chemical potential as described by Eq.~(\ref{eq:disorder}). The index $\ell$ here takes integer values and enumerates the low-energy eigenstates of the numerical BdG Hamiltonian. (a) $V_0=0$ (clean case). (b) $V_0=2\Delta_0$. (c) $V_0=5\Delta_0$. (d) $V_0=10\Delta_0$. We have used $\eta=30$~nm for the disorder correlation length and $N=1000$ for the number of impurities in all panels. The other parameters are the same as in Fig.~\ref{fig:spectrum}.}
		\label{fig:disorder}
	\end{figure*}

\appendix
\section{Current operator in the discretized model}
\label{app:A}
	
In this Appendix, we give the current density operator for the discretized model defined in Eq.~(\ref{eq:H_tb}). At the center of the junction at $y=0$, the current density along the $x$ direction is zero. We therefore focus on the current density operator along the $y$ direction, which can be written as~\cite{Fujimoto2021}
\begin{align}
\bar{j}_y(\boldsymbol{r})&=\frac{e}{\hbar}\Big[\frac{-it}{2 a} \boldsymbol{c}_{n,m}^\dagger\tau_z\sigma_z\left(\boldsymbol{c}_{n,m+1}-\boldsymbol{c}_{n,m-1}\right)\nonumber\\&\quad\quad\ +\frac{\alpha}{4 a}\boldsymbol{c}_{n,m}^\dagger\sigma_y\left(\boldsymbol{c}_{n,m+1}+\boldsymbol{c}_{n,m-1}\right)\Big]+\mathrm{H.c.}
\end{align}
for $\boldsymbol{r}=(na,ma)$. The first line comes from the Wilson mass term and is negligible at low energies, while the second line corresponds to the actual current operator of the TI surface state.

\section{Disorder in the chemical potential}
\label{app:B}

In this Appendix, we show example CdGM spectra in the presence of spatially correlated disorder in the chemical potential. In particular, we model the disorder potential as
\begin{equation}
V(\boldsymbol{r})=V_0\sum_{p=1}^{N}(-1)^p\,e^{-|\boldsymbol{r}-\boldsymbol{r}_p|^2/\eta^2},\label{eq:disorder}
\end{equation}
where $N$ is the number of impurities in the sample, $V_0$ is an average impurity strength, $\boldsymbol{r}_p=(x_p,y_p)$ is the position of the $p$th impurity, and $\eta$ is the correlation length of the disorder. In Fig.~\ref{fig:disorder}, we have plotted the low-energy CdGM spectrum for a single isolated vortex at $x=0$ for a disorder correlation length of $\eta=30$~nm, an impurity number of $N=1000$, and different impurity strengths $V_0$. The other parameters are the same as in Fig.~\ref{fig:spectrum}. Our results are obtained for a single specific disorder configuration without any impurity averaging, as this is the situation that is relevant for a specific experimental sample. Our goal here is to show a specific example for the spectrum of CdGM states in the presence of disorder, while a more detailed study of disorder effects in the TI vortex-based Majorana platform is left to future work.

Figure~\ref{fig:disorder}(a) shows the spectrum in the absence of disorder, $V_0=0$. This is the same plot that was already shown in Fig.~\ref{fig:spectrum}(a) and is just repeated here for convenience. Next, Fig.~\ref{fig:disorder}(b) shows the CdGM spectrum for a disorder strength $V_0=2\Delta_0$. This spectrum still closely resembles the clean case. Figures~\ref{fig:disorder}(c) and (d) show a more disordered scenario with average impurity strengths $V_0=5\Delta_0$ and $V_0=10\Delta_0$, respectively. We find that, for increasing impurity strengths, the spacing between the low-energy CdGM states is significantly reduced, see also Ref.~\cite{deMendon2023,Rechcinski2024}. The reduced CdGM spacing in Figs.~\ref{fig:disorder}(c) and (d) can be understood from the fact that, in the presence of disorder, the average chemical potential near the vortex deviates from the pristine value, which was set to $\mu=0$ here. Since the spacing between CdGM states decreases rapidly as the chemical potential is detuned from the Dirac point (see also Fig.~\ref{fig:ldos_mu}), disorder therefore tends to push the trivial CdGM states towards zero energy.

Note that while we have chosen to vary the impurity strength $V_0$ here, the effects of the disorder potential also depend on the other disorder parameters $N$ and $\eta$, which we have kept fixed for simplicity. Since the numerical values of these parameters are not known for real experimental systems (and they will be strongly material- and sample-dependent anyway), we do not try to make any quantitatively exact predictions here, but merely emphasize the qualitative trend of a reduced CdGM spacing in the presence of disorder. Nevertheless, our results suggest that the vortex MZMs remain reasonably separated in energy for disorder strengths up to a couple of times the proximity-induced superconducting gap. We note that the disorder-induced shift in average chemical potential near the vortex could in principle be compensated by gating as long as the junction is sufficiently wide. In addition, the effects of localized impurities could be partially mitigated by adjusting the position of the vortex by tuning the global phase offset $\phi_0$ via, e.g., a flux loop. Since translational invariance is broken by the disorder, the CdGM spectrum will depend on the spatial position of the vortex, and moving the vortex to a region where the impurity potential is minimal could help maximize the CdGM spacing in a given sample.

\bibliography{refs.bib}

\end{document}